\documentclass[preprint,showpacs,preprintnumbers,amsmath,amssymb,11pt]{revtex4}

\usepackage{graphicx}
\usepackage{dcolumn}
\usepackage{bm}
\usepackage{amssymb}
\usepackage{color}

\def\bea{\begin{eqnarray}}
\def\ena{\end{eqnarray}}

\begin{document}

\title{Limits on the speed of gravitational waves from pulsar timing}

\author{D.~Baskaran}
\affiliation{School of Physics and Astronomy, Cardiff University,
Cardiff CF24 3AA, UK\footnote[1]{e-mail:
Deepak.Baskaran@astro.cf.ac.uk}}
\author{A.~G.~Polnarev}
\affiliation{ Astronomy Unit, School of Mathematical Sciences
Queen Mary, University of London, London E1 4NS,
UK\footnote[2]{e-mail: A.G.Polnarev@qmul.ac.uk}}
\author{M.~S.~Pshirkov}
\affiliation{Pushchino Radio Astronomy Observatory, Astro Space Center, Lebedev Physical Institute, Pushchino, Russia\footnote[3]{e-mail: Pshirkov@prao.ru}}
\author{K.~A.~Postnov}
\affiliation{Sternberg Astronomical Institute, Moscow University, Moscow 119992, Russia\footnote[4]{e-mail: PK@sai.msu.ru}}

\small

\begin{abstract}
{In this work, analyzing the propagation of electromagnetic waves in the field of gravitational waves, we show the presence and significance of the so called surfing effect for pulsar timing measurements. It is shown that, due to the transverse nature of gravitational waves, the surfing effect leads to enormous pulsar timing residuals if the speed of gravitational waves is smaller than speed of light. This fact allows to place significant constraints on parameter $\epsilon$, which characterizes the relative deviation of the speed of gravitational waves from the speed of light. We show that the existing constraints from pulsar timing measurements already place stringent limits on $\epsilon$ and consequently on the mass of graviton $m_g$. These limits on $m_g$ are three orders of magnitude stronger than the current constraints from Solar System tests. The current constraints also allow to rule out massive gravitons as possible candidates for cold dark matter in galactic halo. In the near future, the gravitational wave background from extragalactic super massive black hole binaries, along with the expected sub-microsecond pulsar timing accuracy, will allow to achieve constrains of $\epsilon\lesssim0.4\%$ and possibly stronger.}

\end{abstract}



\today

\maketitle



\section{Introduction \label{Introduction}}

Gravitational wave astronomy is an active field of research which promises to open up a new window into the physical universe  \cite{Thorne1987}, \cite{Allen1997}, \cite{glpps2001},\cite{CutlerThorne2001}, \cite{Hughes2003}, \cite{Grishchuk2003}, \cite{sathya2005}. The current and future laser interferometric gravitational wave detectors, high precision pulsar timing, along with measurements of the anisotropies in the temperature and polarization of the Cosmic Microwave Background have the potential to discover gravitational waves in a broad range of frequencies in the near future (see \cite{LIGOwebsite}, \cite{Jenetetal2006}, \cite{PLANCKbluebook} for recent discussions).

In this paper we shall be mainly interested in pulsar timing as a laboratory for gravitational wave physics. Propagation of pulsar signal through space-time perturbed by gravitational waves results in appearance of anomalous timing residuals (i.e.~differences between observed  and theoretically predicted times of arrival). Pulsar timing  provides a unique tool for observing gravitational waves in low-frequency band ($10^{-7} ~\mathrm{Hz}<f_{gw}<10^{-9}  ~\mathrm{Hz} $) \cite{Sazhin1978}, \cite{Detweiler1979}, \cite{Bertotti1983}, \cite{Cordes2004}, \cite{Hobbs2005}, \cite{Jenetetal2005}, \cite{Jenetetal2006}. The main sources of gravitational waves at these frequencies are expected to be of extragalactic origin. The strongest sources would be supermassive black hole binaries in the center of galaxies \cite{WyitheLoeb2003}, \cite{JaffeBacker2003}, \cite{Enoki2004}, \cite{Sesana2008}. Relic gravitational waves, which are the remnants from the early history of the universe, may also contribute a significant fraction to the gravitational wave background at these frequencies \cite{Grishchuk1974}, \cite{Grishchuk2005}. Pulsar timing could also measure gravitational waves from superstrings \cite{Maggiore2000}, as well as several other exotic sources \cite{Hogan2006}.

The main methods to detect gravitational waves are based on the analysis of their interaction with electromagnetic fields \cite{LandauLifshitz}, \cite{mtw}, \cite{GrishchukPolnarev1980}. The interaction of gravitational waves with electromagnetic waves leaves measurable imprints on the latter. For example, the phase variations in the electromagnetic wave propagating in the field of a gravitational wave, and its implications for space radio interferometry were studied in \cite{bkpn1990} (see also \cite{bkpn1992}). In  \cite{PolnarevBaskaran2008}, analyzing these phase variations in a situation when the speed of gravitational waves could be smaller than the speed of light, the authors introduced the concept of ``surfing effect" and studied its implications for the precision interferometry measurements. In this paper we shall consider the implications of the surfing effect for pulsar timing measurements. As we shall show, due to the transverse nature of gravitational waves, the surfing effect can lead to enormous observable pulsar timing residuals if the speed of gravitational waves is smaller than the speed of electromagnetic waves.  We shall use this fact, along with the expected precision of pulsar timing measurements, to place stringent upper limits on the parameter $\epsilon=(c-v_{gw})/c$ which characterizes the deviation of speed of gravitational waves from the speed of light. We show that, for a realistic gravitational wave background and a reasonable time duration of observations, the achievable limits are $\epsilon\lesssim 0.4 \%$. Constraining the speed of gravitational waves is an interesting experimental challenge attracting much theoretical and experimental interest \cite{WillBook}, \cite{Will2001}, \cite{Kopeikin2004}. We argue that the constraint on $\epsilon$ from pulsar timing would provide the strongest current limitations on the deviation of speed of gravitational waves from speed of light.

It is worth mentioning that the surfing effect considered in this paper is quite generic. The surfing effect occurs in any physical situation where the phase speed of gravitational waves is smaller than the phase speed of electromagnetic waves \cite{bkpn1990}, \cite{PolnarevBaskaran2008}. For example, this is the case in theories which predict a non vanishing rest mass for graviton \cite{MassiveGravity}, \cite{WillBook}, \cite{BabakGrishchuk2003}. Although, generically, these theories predict extra polarization states for gravitational waves, in our work we shall restrict our analysis to effects caused only by transverse traceless (TT) gravitational waves. Another possible scenario for the surfing effect to arise is to consider the interaction of gravitational waves and electromagnetic waves in the presence of plasma. In this case the phase speed of gravitational waves remains unchanged and is equal to $c$ (i.e.~the speed of light in vacuum), while the phase speed of electromagnetic waves becomes generally greater than $c$ \cite{Jackson}.

The plan of the paper is as follows. We shall begin in Section \ref{SingleWave} with the analysis of propagation of an electromagnetic wave in the field of a single monochromatic plane gravitational wave. We shall calculate the timing residuals due to a single gravitational wave and discuss the manifestations and physical consequences of the surfing effect. In Section \ref{ArbitraryWaveField} we generalize the surfing effect for the case of an arbitrary gravitational wave field. We derive the statistical properties of the timing residual signal based on the statistical properties of the gravitational wave field. In Section \ref{upperlimits} we calculate the achievable constraints on $\epsilon$ depending on the strength of the gravitational wave background characterized by energy density parameter $\Omega_{gw}$. In Section \ref{physicalconsequences} we study the physical consequences of the surfing effect in pulsar timing. We show that the gravitational wave background from extragalactic black holes allows to place strong limits on $\epsilon$. Furthermore we show that the surfing effect can also place a strong upper bound on the mass of graviton. Finally, we conclude the paper in Section \ref{conclusions} with a summary of the main results of this work.


\section{Pulsar timing residuals for a single monochromatic gravitational wave\label{SingleWave}}

In this paper we shall be working in the framework of a slightly perturbed Minkowski space time with coordinates $x^\mu = (ct,x^i)$ and the metric given by
\bea
ds^2 = -c^2dt^2+\left(\delta_{ij}+h_{ij}\right)dx^idx^j,
\label{metric}
\ena
where $h_{ij}$ is the gravitational wave perturbation. For clarity and in order to gain physical insight into the problem, in this section, we shall consider the case of a single monochromatic plane gravitational wave. In the next section, we shall generalize our analysis to the case of an arbitrary gravitational wave field. For a monochromatic gravitational wave the metric perturbation $h_{ij}$ takes the form \cite{LandauLifshitz}, \cite{mtw}
\bea
h_{ij} = h~p_{ij}e^{ik_\mu x^\mu} = h~p_{ij}e^{-i\left(k_0ct-k_ix^i\right)},
\label{singlegwmetric}
\ena
where $h$ is the amplitude of the gravitational wave, $k_\mu = \left(k_0,k_i\right)$ is the wave vector, and $p_{ik}$ is the polarization tensor of the gravitational wave.  Introducing a set of two mutually orthogonal unit vectors $l_i$ and $m_i$ orthogonal to the wave vector $k_i$, the polarization tensor $p_{ik}$ has the form \cite{LandauLifshitz}, \cite{mtw}
\bea
p_{ik} = \frac{1}{2}\left( l_i\pm m_i \right)\left( l_k\pm m_k \right),
\label{defpolten}
\ena
where $\pm$ corresponds to the two independent states of circular polarization. Due to the transverse and traceless nature of gravitational waves, the polarization tensor satisfies the following conditions
\bea
p_{ik}k^i = 0, ~~~p_{ik}\delta^{ik} = 0.
\label{TTconditions}
\ena
For further discussion, it is convenient to introduce the wavenumber $k=\left(\delta_{ij}k^ik^j\right)^{1/2}$, and a unit vector in the direction of wave propagation $\tilde{k}^i=k^i/k$. The wavelength of the gravitational wave is related to the wavenumber by the equality $k = 2\pi/\lambda_{gw}$. The frequency of the gravitational wave ${f_{gw}}$ is related to the time component of the wave vector through the relation $k_0=2{\pi}{f_{gw}}/c$.

The speed of a gravitational wave is determined by relationship $v_{gw}= {{f_{gw}}}\lambda_{gw}$. In General Relativity gravitational waves travel at the speed of light, i.e.~$v_{gw}=c$, which implies a relationship (dispersion relationship) $k = k_0$. In order to analyze the possibility $v_{gw}\neq c$, let us introduce a phenomenological parameter $\epsilon$ describing the relative deviation of $v_{gw}$ from speed of light $c$
\bea
\epsilon \equiv \frac{c-v_{gw}}{c}, ~~~\textrm{where}~~~v_{gw}
\equiv {{f_{gw}}}\lambda_{gw} = \frac{ck_0}{k} = c\left( 1-\epsilon \right).
\label{epsilon}
\ena
The quantity $\epsilon$ has been introduced as a phenomenological parameter, and thus the analysis that follows is valid for any theory that predicts gravitational waves with $v_{gw}\neq c$. Particularly, of interest are modifications of General Relativity that predict massive gravitons. For these models, $\epsilon$ can be related to the rest mass of the graviton $m_g$ through the relation
\bea
\epsilon = 1 - \frac{\hbar ck_0}{\hbar ck_0 + m_gc^2} \approx \frac{m_gc}{\hbar k_0} = \frac{m_gc^2}{2\pi\hbar {{f_{gw}}}}.
\label{mgravitondef}
\ena

Let us move our attention to pulsar timing measurements. The effect of a gravitational wave upon the measured frequency of pulsar signal is given by \cite{Sazhin1978}, \cite{Detweiler1979},
\bea
\frac{\Delta\nu(t)}{\nu_0} = \frac{1}{2c}\int\limits_0^D ds\left.~ \left(e^ie^j\frac{\partial h_{ij}}{\partial t}\right)\right|_{path},
\label{deltanu1}
\ena
where $\nu_0$ is the unperturbed pulsar frequency in the absence of gravitational waves and $\Delta\nu(t) = \nu(t) - \nu_0$ is the variation of pulsar frequency due to the presence of a gravitational wave. $D$ is the distance from the pulsar to the observer, integration variable $s$ is the distance parameter along the unperturbed light ray path from pulsar to the observer, $e^i$ is the unit vector tangent along this path (i.e.~unit vector in the direction from pulsar to the observer), and the subscript indicates the integration along this path. The unperturbed light ray path is given by
\bea
t(s) = t - \frac{s}{c}, ~~~~ x^i(s) = x^i - e^is,
\label{lightraypath}
\ena
where $t$ and $x^i$ determine the time and position of the observation. Without loss of generality we can set $x^i=0$ by choosing a spatial coordinate system with observer at its origin.

Substituting the path (\ref{lightraypath}) into (\ref{deltanu1}), taking into account (\ref{singlegwmetric}) and (\ref{epsilon}), after straight forward integration we arrive at
\bea
\frac{\Delta\nu(t)}{\nu_0} = \frac{1}{2}\left(1-\epsilon\right)h~e^ie^jp_{ij}~e^{-ik(1-\epsilon)ct}~\left[ \frac{1-e^{i\left(1-\epsilon-\tilde{k}_ie^i\right)kD}}{\left(1-\epsilon-\tilde{k}_ie^i\right)} \right].
\label{deltanu2}
\ena

The pulsar timing measurements customarily measure the timing residuals, i.e.~the difference between the actual pulse arrival times and times predicted from a spin-down model for a pulsar \cite{Detweiler1979}, \cite{Hobbs2005}. The variations in the measured frequency, due to the presence of a gravitational wave, will cause an anomalous timing residual $R(t)$ in the pulse arrival time given by \cite{Detweiler1979}
\bea
R(t) = \int\limits_{t-T}^t dt ~ \frac{\Delta\nu(t)}{\nu_0},
\label{residual1}
\ena
where $T$ is the time of observations, and the residual $R(t)$ is measured in seconds. Substituting expression (\ref{deltanu2}) into (\ref{residual1}), we get for the timing residual due to a single monochromatic gravitational wave
 \bea
 R(t) = \frac{i}{2kc}h~e^ie^jp_{ij}~e^{-ik\left(1-\epsilon\right)ct}~\left(1-e^{ik\left(1-\epsilon\right)cT}\right)~\left[ \frac{1-e^{i\left(1-\epsilon-\tilde{k}_ie^i\right)kD}}{\left(1-\epsilon-\tilde{k}_ie^i\right)}\right].
\label{residual2}
 \ena
Before proceeding further, let us analyze the above expression. The expression in the square brackets on the right side of (\ref{residual2}) becomes large (proportional to $kD\sim D/\lambda_{gw}$) when $\left(1-\epsilon-\tilde{k}_ie^i\right)\rightarrow 0$, i.~e. 
\bea
R(t) \approx  \frac{1}{2kc}h~e^ie^jp_{ij}~e^{-ik\left(1-\epsilon\right)ct}\left(1-e^{ik\left(1-\epsilon\right)cT}\right)\left[ kD\left( 1+ \frac{}{}O\left(\delta\right)\right)\right], ~ {\rm for} ~\delta\equiv\left(1-\epsilon-\tilde{k}_ie^i\right)kD \ll 1.
\label{residual2a}
\ena
Hence, for gravitational waves traveling in a direction at a sufficiently small angle to the direction from the pulsar, i.e.~$\tilde{k}_ie^i \approx \left(1-\epsilon\right)$, there is a resonance inrease in the expression for timing residual. In the case when $\epsilon =0$ this does not lead to a growth of the timing residual $R(t)$ itself, due to the transverse nature of the gravitational wave (since $e^ie^jp_{ij}\rightarrow 0$ when $\tilde{k}_ie^i\rightarrow1$, see expression (\ref{eep})). On the other hand, if $\epsilon\neq0$, the expression for $R(t)$ increases significantly for $\tilde{k}_ie^i \approx  \left(1-\epsilon\right)$. The resonance occurs when the signal from the pulsar ``surfs" along the gravitational wave, i.e.~travels at a small angle $\cos{\theta}\approx  \left(1-\epsilon\right)$ to the gravitational wave. This picture is reminiscent of wave surfing, so for this reason following \cite{PolnarevBaskaran2008} we call this effect, of a resonant increase in $R(t)$, as the surfing effect. It is worth noticing that the above analysis closely resembles considerations in \cite{PolnarevBaskaran2008}, where the surfing effect manifested itself in the resonance growth of the phase variation of electromagnetic waves, leading to an observable angular displacement of distant quasars. In the current work, we are analyzing the signature of the surfing effect in pulsar timing residuals.


\section{Pulsar timing residuals for an arbitrary gravitational wave field\label{ArbitraryWaveField}}

In the previous section we calculated the timing residual due to a single plane monochromatic gravitational wave. In this section we shall generalize our analysis to an arbitrary gravitational wave field. In general, an arbitrary gravitational wave field can be decomposed into spatial Fourier modes
\bea
h_{ij}(t,x^i) = \int d^3{\bf{k}} \sum_{s=1,2} \left[
h_s(k^i,t)\stackrel{s}{p}_{ij}(k^l)e^{ik_ix^i} +
h_s^*(k^i,t)\stackrel{s}{p}_{ij}^*(k^l)e^{-ik_ix^i}
\right],
\label{fouriergw}
\ena
where $d^3{\bf k}$ denotes the integration over all possible wave vectors, and $s=1,2$ corresponds to the two linearly independent modes of polarization satisfying the orthogonality condition
\bea
\stackrel{s}{p}_{ij}\stackrel{s'}{p}{}^{ij*} = \delta_{ss'}
\label{poltenorthog}
\ena
The mode function $h_s(k^i,t)$ correspond to plane monochromatic waves
\bea
h_s(k^i,t) = h_s(k^i)~e^{-ik\left(1-\epsilon\right)ct}
\label{hmodefunctions}
\ena
Due to the linear nature of the problem, following the decomposition (\ref{fouriergw}), the total timing residual due to an arbitrary gravitational wave field, can be presented in the following manner
\bea
R(t) = \int d^3{\bf{k}} \sum_{s=1,2} \left[
h_s(k^i)\tilde{R}(t;k^i,s) +
h_s^*(k^i)\tilde{R}^*(t;k^i,s)
\right].
\label{fourierR}
\ena
Using the results of the previous section, the contribution from a single Fourier component $\tilde{R}(t;k^i,s)$ is given by
\bea
\tilde{R}(t;k^i,s) = \frac{i}{2kc}~e^ie^jp_{ij}~e^{-ik\left(1-\epsilon\right)ct}~\left(1-e^{ik\left(1-\epsilon\right)cT}\right)~\left[ \frac{1-e^{i\left(1-\epsilon-\tilde{k}_ie^i\right)kD}}{\left(1-\epsilon-\tilde{k}_ie^i\right)}\right],
\label{residual3}
\ena
where the tilde over $R$ in the above expression is introduced to indicate explicit factoring out of the gravitational wave amplitude $h$ compared with (\ref{residual2}).

In general, if we have the information about the mode functions $h_s(k^i)$, using expressions (\ref{fourierR}) and (\ref{residual3}) we can calculate the expected timing residual for an arbitrary gravitational wave field. In most of the practically interesting cases we do not have such a complete knowledge of the gravitational wave field, but are restricted to the knowledge of its statistical properties. To proceed, let us assume the following statistical properties
\bea
<h_s(k^i)> = 0,~~~ <h_s(k^i)~ h_{s'}^{*}(k'^i)> =
\frac{P_h(k)}{16\pi k^3}\delta_{ss'}\delta^3(k^i-k'^i),
\label{gwstatprop}
\ena
where the brackets denote ensemble averaging over all possible realizations, and $P_h(k)$ is the metric power spectrum per logarithmic interval of $k$. These conditions correspond to a stationary statistically homogeneous and isotropic gravitational wave field.

The positing of the statistical properties of the gravitational wave field (\ref{gwstatprop}) allows us to calculate the statistical properties of the timing residual $R(t)$. Using (\ref{fourierR}) and (\ref{gwstatprop}), and taking into account the orthogonality property (\ref{poltenorthog}), after straight forward calculations, we arrive at the following statistical properties for the timing residual $R$
\begin{subequations}
\bea
<R(t)> &=& 0,
\label{Rmean}\\
 <R^2(t)>
&=& \int \frac{dk}{k} P_h(k) \tilde{R}^2(k),
\label{Rsquaremean}
\ena
\end{subequations} where we have introduced the transfer function
\bea
\tilde{R}^2(k) = \frac{1}{8\pi}\int d\Omega \sum_s
\left| \tilde{R}(t;k^i,s) \right|^2.
\label{transferfunction}
\ena
In the above expression $d\Omega$ represents integration over the possible directions of gravitational wave (i.e.~$d^3{\bf k} = k^2dkd\Omega$). From (\ref{residual3}) and (\ref{transferfunction}) it follows that the transfer function $\tilde{R}^2(k)$ does not depend on time variable $t$, which is a reflection of the stationarity of the underlying gravitational wave field.

\begin{figure}
\begin{center}
\includegraphics[width=7cm]{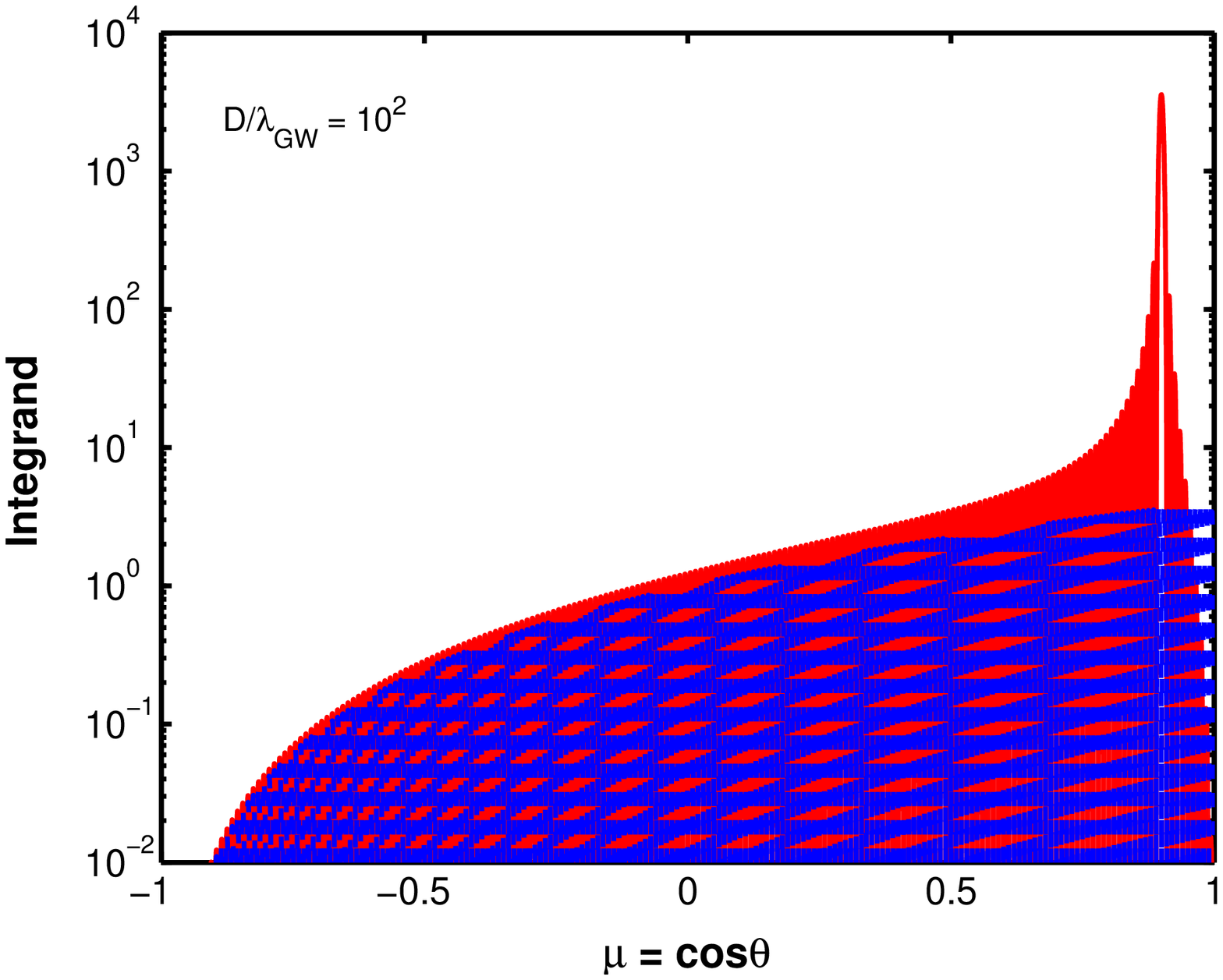}
\includegraphics[width=7cm]{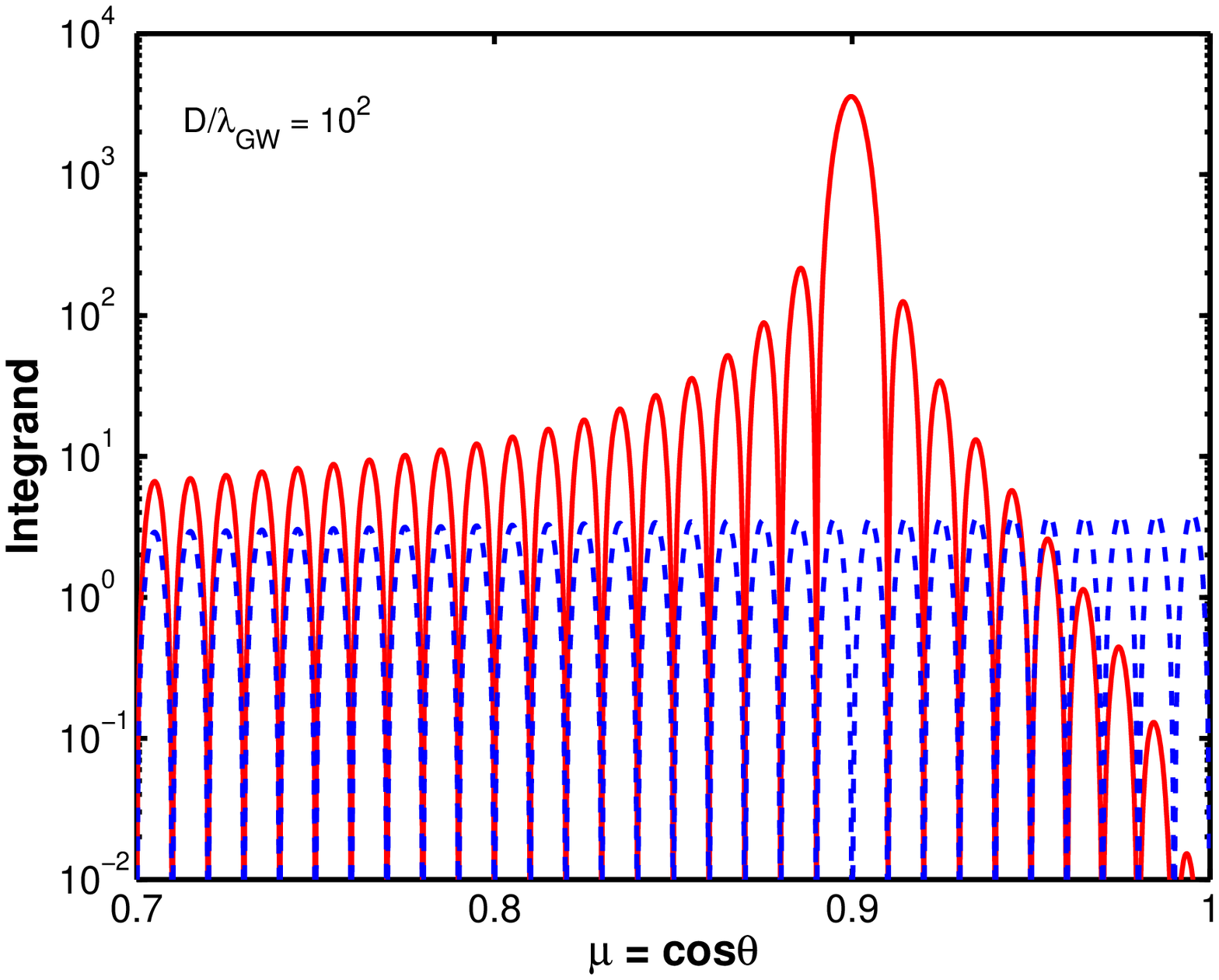}
\end{center}
\caption{The illustration of the resonance effect, present for
$\epsilon \neq 0$. The graphs show integrand in expression
(\ref{transferfunction2}). For the case $\epsilon\neq 0$ the
integrand sharply peaks at angle $\mu\approx (1-\epsilon)$ (solid
red line), while for the case $\epsilon = 0$ the effect is absent
(dashed blue line). In the case of $\epsilon \neq 0$, the
gravitational waves travelling at an angle $\cos{\theta} \approx
(1-\epsilon)$ to the line of sight are the predominant
contributors to the surfing effect. The figure on the left shows
the integrand for the whole region of $\mu$, while the figure on
the right zooms into the region around the
resonance.}\label{figure1}
\end{figure}

The expression for the transfer function can be explicitly calculated. In order to do this, let us firstly introduce a spherical coordinate system $(\theta,\phi)$ related to the spatial coordinates $\{x^i\}$ (following notations of \cite{Goldstein}). Without loss of generality, we can assume that our spatial coordinate system is chosen such that the unit vector from the pulsar to the observer points in the north-pole direction, i.e.~$e^i = (0,0,1)$. Let us also introduce the quantity $\mu = \cos{\theta} = e_i\tilde{k}^i$,
characterizing the angle between the direction of gravitational wave propagation and the direction from pulsar to the observer. Furthermore, let $\phi$ denote the azimuthal angle that is subtended by $\tilde{k}^i$ projected onto the $(x^1,x^2)$-plane, i.e.~$\tilde{k}^1=\cos{\phi}\sin{\theta}$ and $\tilde{k}^2=\sin{\phi}\sin{\theta}$. Introducing $e^{\theta}_i$ and $e^{\phi}_i$ which are the meridian and azimuthal unit vectors perpendicular to the gravitational wave wavevector $k_i$ respectively, the polarization tensors for gravitational waves (\ref{defpolten}) take the form $\stackrel{s}{p}_{ij}(k^i) = (e^{\theta}_i\pm i e^{\phi}_i)(e^{\theta}_j\pm i e^{\phi}_j)/2$, with $\pm$ corresponding to the two independent circularly polarized degrees of freedom $s=1,2$ (for a detailed discussion see for example \cite{dgp2006}, \cite{Baskaran2004}). Taking into account the relation
\bea
e^ie^j\stackrel{s}{p}_{ij} = \frac{1}{2}(1-\mu^2)e^{\pm
2i\phi},
\label{eep}
\ena
substituting (\ref{residual3}) into (\ref{transferfunction}) and setting $d\Omega = d\mu d\phi$, after integration over $\phi$, we arrive at the expression for the transfer function
\bea
\tilde{R}^2(k) = \frac{1}{2k^2c^2}\sin^2\left(\frac{kcT}{2}\left(1-\epsilon\right)\right)~\int\limits_{-1}^{+1} d\mu \left(1-\mu^2\right)^2 \left[ \frac{\sin^2\left\{\frac{kD}{2}\left(1-\epsilon-\mu\right) \right\}}{\left(1-\epsilon-\mu\right)^2} \right].
\label{transferfunction2}
\ena
The integrand under the integral in the above expression is illustrated in Figure \ref{figure1}. As can been seen, when $\epsilon\neq 0$, the predominant contribution to the integral comes from the resonance region $\mu \approx \left(1 - \epsilon\right)$. Thus, in this case, the predominant contribution to the timing residual $<R^2>$ comes from ``surfing" gravitational waves, i.e.~waves for which $\mu \approx \left(1 - \epsilon\right)$. In the physically interesting limit $\epsilon\rightarrow 0$ and $kD\rightarrow \infty$ we can calculate the integral in (\ref{transferfunction2}) explicitly. We refer the reader to Appendix \ref{AppendixA} for details of this calculation. The result is as follows
\bea
\tilde{R}^2(k) \approx \frac{2}{3k^2c^2}\sin^2\left(\frac{kcT}{2}\left(1-\epsilon\right)\right)~\ \left[  1 + \frac{3}{2}\pi\epsilon^2kD \right].
\label{transferfunction3}
\ena
The above expression allows us to simply quantify the condition for the surfing effect to be dominant, $\epsilon^2kD \gg1$. As we shall show in the next section, given the precision level of the current and planned pulsar timing measurements, the surfing effect allows to place significant constraints on the $\epsilon$ parameter.

Before proceeding, it is instructive to compare the results of this section with the results of \cite{PolnarevBaskaran2008}. More specifically, it is interesting to compare expression (\ref{transferfunction3}) for the transfer function of timing residuals with its counterpart expression (29) in \cite{PolnarevBaskaran2008} for the transfer function of angular displacement
\bea
\Delta\tilde{\alpha}^2(k) \approx \frac{1}{20}\left[ 1 + 5\pi\epsilon^3kD \right].
\nonumber
\ena
Apart from the differing factors in front of the square brackets in the two expression, the crucial difference is the differing powers of $\epsilon$. In the present work, the surfing effect manifests in the term $\epsilon^2kD$ in the square brackets of (\ref{transferfunction3}). In \cite{PolnarevBaskaran2008}, the surfing effect manifests in the term $\epsilon^3kD$ term in the square brackets of (29). The extra factor of $\epsilon$ arose due to the geometrical specificity of interferometric observations of phase difference at the ends the interferometric system (see \cite{PolnarevBaskaran2008} for details). The main consequences of this difference are twofold. Firstly, equivalent constraints on $\epsilon$ require smaller distance to the source in the case of pulsar timing compared with interferometric observations. This is reflected in the fact that in the present work we focus on galactic pulsars, where as \cite{PolnarevBaskaran2008} focused on high redshift quasars. Secondly, the condition for surfing effect to dominate is different in the two contexts. This condition, characterized by the value of $\epsilon_*$ (see expression (\ref{epsilonstar}) below and expression (32) in \cite{PolnarevBaskaran2008}), places the lower limit on the potentially possible bounds on $\epsilon$. This limiting bound is lower for interferometry measurements ($\epsilon_*\approx 2.3\times10^{-4}$) than for pulsar timing measurements ($\epsilon_*\approx3.2\times10^{-3}$). Even so, due to exceptional precision, the experimentally achievable bounds on $\epsilon$ from pulsar timing measurements would be more stringent. 

\section{Upper limits on the speed of gravitational waves\label{upperlimits}}

Let us now turn our attention to the various cosmological and astrophysical candidates for a stochastic gravitational wave background and their contribution to the surfing effect in pulsar timing measurement. Analyzing their magnitude, we shall study the achievable upper limits on $\epsilon$ that these backgrounds could place.

The stochastic gravitational wave field may be characterized by the dimensionless strain amplitude $h_c(f)$ which is related to the power spectrum $P_h$ in the following way
\bea
h_c(f) \equiv \sqrt{P_h(k)}, ~~~\mathrm{where} ~ f = \frac{c k}{2\pi}(1-\epsilon).
\label{definitionhc}
\ena
The quantity $h_c(f)$ is the root-mean value of the gravitational wave amplitude in a unit logarithmic interval of frequencies. For analyzing the stochastic gravitational wave fields, it is also customary to introduce the density parameter $\Omega_{gw}$ to characterize the strength of the gravitational wave field \cite{Allen1997}, \cite{glpps2001}, \cite{Grishchuk2003}. $\Omega_{gw}$ is related to the power spectrum $P_h(k)$ and strain $h_c(f)$ by the relation
\bea
\Omega_{gw}(k)  =   \frac{2\pi^2}{3} \left( \frac{k}{k_H}\right)^2P_h(k)  
 =  \frac{2\pi^2}{3} \left( \frac{f}{f_H}\right)^2 h_c^2(f) \label{definitionofOmega}
\ena
where $k_H= 2\pi f_H/c = 2\pi H_o/c$, and $H_o$ is the current Hubble parameter. The density parameter $\Omega_{gw}$ is the current day ratio of energy density of gravitational waves (per unit logarithmic interval in $k$) to the critical density of the Universe $\rho_{crit} = 3c^2H_o^2/8\pi G$. Below, for numerical estimations, we set Hubble parameter $H_o = 75~\frac{{\rm km}}{{\rm sec}}/{\rm Mpc}$. Note, that the above definition (\ref{definitionofOmega}) is valid for stationary gravitational wave backgrounds. In cosmological context, when considering relic gravitational, waves this definition modifies to $\Omega_{gw}(k)  =   \frac{\pi^2}{3} \left( \frac{k}{k_H}\right)^2P_h(k)$ due to the non-stationary (standing wave) nature of relic gravitational waves (see for example \cite{glpps2001}).

For simplicity, in the numerical estimations below, we shall assume a simple power law behaviour for $h_c$ which is equivalent to a power law spectrum for the density parameter $\Omega_{gw}$
\bea
h_c(f) = h_c(f_o)\left(\frac{f}{f_o}\right)^{\alpha}, ~~~~ \Omega_{gw} (k)= \Omega_{gw} (k_o) \left(\frac{k}{k_o}\right)^{n_T},
\label{powerlawspectrum}
\ena
where
\bea
\Omega_{gw}(k_o) = \frac{2\pi^2}{3}\left(\frac{k_o}{k_H}\right)^2h_c^2(f_o),~~~~k_o = \frac{2{\pi}f_o}{c}, ~~~ n_T = 2(1+\alpha).
\label{powerlawspectrum2}
\ena
Although restricted, this form of spectrum is a good approximation for a large variety of models in gravitational wave frequency range of our interest. For example, this type of power law spectrum, with $\alpha = -2/3$, is produced by the extragalactic coalescing super massive binary black hole systems \cite{WyitheLoeb2003}. In cosmological context, this type of a power spectrum, with spectral index $\alpha$ at the current epoch, arises due to the evolution of relic gravitational waves with a primordial spectral index equal to $2(1+\alpha)$, (i.e.~$P_h(k)|_{prim}\propto k^{2(1+\alpha)}$) \cite{Grishchuk1974}. The flat, scale invariant power spectrum (also known as Harrison-Zeldovich power spectrum) corresponds to $\alpha=-1$ (i.e.~$n_T=0$). In general the power law spectrum just assumes the absence of features in the spectrum of gravitational waves at the wavelengths of our interest.

In practice, when considering pulsar timing,  we are interested in calculating the expected mean square deviation of the timing residuals due to stochastic background of gravitational waves (\ref{Rsquaremean}). In order to evaluate $<R^2(t)>$ from expression (\ref{Rsquaremean}) we require to specify the limits of integration $k_{min}$ and $k_{max}$. $k_{min}$ and $k_{max}$
determine the frequency range of gravitational waves that can be probed by pulsar timing measurements. The lower limit $k_{min}$ is determined by the time duration of observations $T_{obs}$, $k_{min} \approx 2{\pi}f_{obs}/c = 2{\pi}/cT_{obs}$. In our estimates we shall assume $T_{obs} \approx 10~{\rm yrs}$. The upper limit $k_{max}\approx 2\pi/c\delta t$ is determined by the duration of single observation $\delta t$ (in other words, the time of integration), which is usually of the order of 1-2 hours. We note here that it is this time (and not the time between consecutive observations, of order of weeks) which determines $k_{max}$ in timing residuals. Indeed, if the period of a gravitational wave is smaller than $\delta t$, its effect is smeared out by averaging procedure. But if the period of the gravitational wave lies between the averaging time and the sampling time, the wave will clearly manifest itseft in the timing residuals. Some authors erroneously use the inverse sampling time as $k_{max}$, apparently guided by the analogy with time series analysis. Thus, in our case, it is safe to assume $\delta t\ll T_{obs}$ (i.e.~$k_{max}\gg k_{min}$), and set $k_{max} = \infty$ in numerical evaluations below. Furthermore, we shall be working under the assumption $kD = 2\pi D/\lambda_{gw} \gg 1$, which corresponds to the reasonable assumption that the gravitational waves of our interest ($\lambda_{gw}\lesssim 10~l{\mathrm{yrs}}$) have wavelengths much shorter than the distance to the pulsar ($D\sim 10~{\mathrm{kpc}}$).

As can be seen from expression (\ref{transferfunction3}) and the considerations in Appendix \ref{AppendixA}, the
behaviour of the transfer function $\tilde{R}^2(k)$
depends on value of the quantity $3\pi\epsilon^2kD/2$. In order to
analyze the various possibilities let us introduce
\bea
\epsilon_* = \left(\frac{3}{2}\pi k_{min}D\right)^{-1/2} = 3.2 {\times}
10^{-3}\left[\left( \frac{10~{\mathrm{kpc}}}{D} \right)\left( \frac{T_{obs}}{ 10~{\mathrm{yrs}} } \right)\right]^{\frac{1}{2}}.
\label{epsilonstar}
\ena
Below we shall analyze the two possibilities, $\epsilon\ll\epsilon_*$ and $\epsilon\gg \epsilon_*$, separately.

In the case $\epsilon\ll\epsilon_*$ in transfer function $\tilde{R}^2(k)$, in expression (\ref{transferfunction3}), we can neglect the second term in the square brackets in comparison with the first. Furthermore, in the term $\sin^2{\left(kcT\left(1-\epsilon\right)\right)/2}$ we can neglect the rapid oscillatory factor. Thus, for the transfer function we get
\bea
\tilde{R}^2(k) \approx \frac{2}{3k^2c^2}\left(\frac{1-\cos\left({kcT}\left(1-\epsilon\right)\right)}{2}\right) \left[  1 + \frac{3}{2}\pi\epsilon^2kD \right]  \approx  \frac{1}{3k^2c^2}.
\label{transferfunctionepsilonllepsilonstar}
\ena
Substituting the above approximation (\ref{transferfunctionepsilonllepsilonstar}), taking into account the definition (\ref{definitionhc}) and a power law spectrum (\ref{powerlawspectrum}), into expression (\ref{Rsquaremean}), and setting the limits of integration as mentioned above, we arrive at
\bea
<R^2(t)>  \approx \frac{T_{obs}^2h_c^2(f_{obs})}{24\pi^2\left(1-\alpha\right)}, ~~~~\textrm{for}~\epsilon\ll\epsilon_*.
\label{Rsquare1}
\ena

In the case $\epsilon\gg\epsilon_*$, neglecting the first term in the square brackets with respect to the second in (\ref{transferfunction3}) and ignoring the rapid oscillatory factor, the transfer function can be approximated as
\bea
\tilde{R}^2(k) \approx \frac{2}{3k^2c^2}\left(\frac{1-\cos\left({kcT}\left(1-\epsilon\right)\right)}{2}\right) \left[  1 + \frac{3}{2}\pi\epsilon^2kD \right]  \approx  \frac{\pi\epsilon^2D}{2kc^2}.
\label{transferfunctionepsilonggepsilonstar}
\ena
In this case, the expression for (\ref{Rsquaremean}) takes the form
\bea
<R^2(t)>  \approx \frac{T_{obs}^2h_c^2(f_{obs})}{12\pi^2\left(1-2\alpha\right)}\left(\frac{\epsilon}{\epsilon_*}\right)^2, ~~~~\textrm{for}~\epsilon\gg\epsilon_*.
\label{Rsquare2}
\ena
Comparing expressions (\ref{Rsquare1}) and (\ref{Rsquare2}) it can be seen that when $\epsilon\gg\epsilon_*$ the surfing effect leads to a strong resonance contribution (proportional to $kD$) in the timing residual compared with the case when $\epsilon\ll\epsilon_*$. This dominant resonance contribution comes from gravitational waves traveling at an angle $\cos{\theta}\approx\left(1-\epsilon\right)$ to the direction of signal propagation from the pulsar (see Appendix \ref{AppendixA} for details).

\begin{figure}
\begin{center}
\includegraphics[width=12cm]{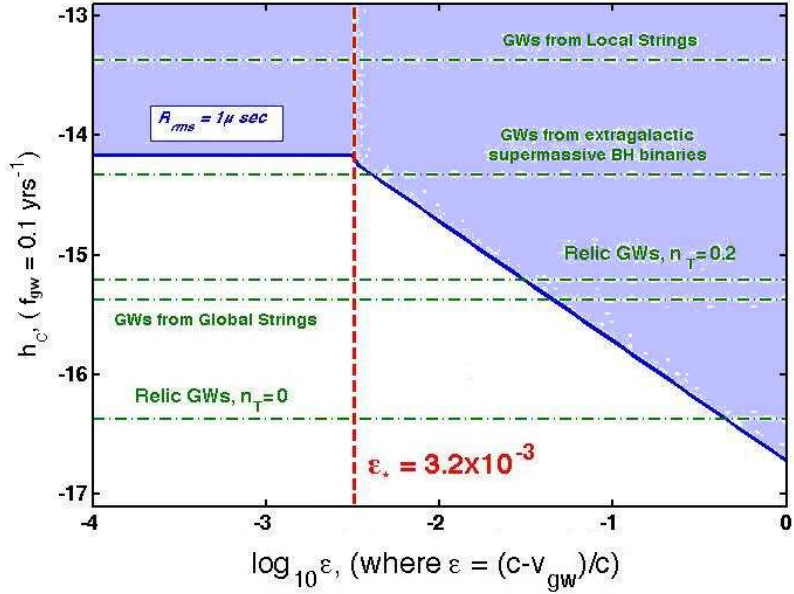}
\end{center}
\caption{The upper limit on strain amplitude $h_c$ and
velocity parameter $\epsilon$ for gravitational waves, achievable by pulsar timing residual measurements with precision $R_{rms}=0.1~{{\mu}\mathrm{sec}}$ and time of observation $T_{obs} = 10~{\mathrm{yrs}}$. The shaded area shows the region that can be probed or ruled out by pulsar timing observations. The horizontal lines show the strain $h_c$, at $f = (10~\mathrm{{\mathrm{yrs}}})^{-1}$, for some viable sources of gravitational wave.}\label{figure2}
\end{figure}

From expressions (\ref{Rsquare1}) and (\ref{Rsquare2}), it follows that the direct measurement of pulsar timing residuals would be able to measure or constrain either $h_c$ or $h_c\epsilon$, depending on the value of $\epsilon$ compared with $\epsilon_*$. A null result in timing residual measurements would place the following upper limits
\bea
h_c \lesssim 4.9\times10^{-15}\left[ \sqrt{1-\alpha}\left(\frac{R_{rms}}{0.1~{\mu}{\rm sec}}\right)\left(\frac{10~{\rm yrs}}{T_{obs}}\right) \right], ~~~~\textrm{for}~\epsilon\ll\epsilon_*,
\label{hc_limit}
\ena
or
\bea
h_c\epsilon \lesssim 1.1\times 10^{-17}\left[ \sqrt{1-2\alpha}\left(\frac{R_{rms}}{0.1~{\mu}{\rm sec}}\right)\left(\frac{10~{\rm yrs}}{T_{obs}}\right)^{\frac{1}{2}} \left(\frac{10~{\mathrm{kpc}}}{D}\right)^{\frac{1}{2}} \right], ~~~~\textrm{for}~\epsilon\gg\epsilon_*.
\label{hcepsilon_limit}
\ena
where $R_{rms} = \sqrt{<R^2(t)>}$ is the precision of the pulsar residual timing, and $h_c=h_c(f_{obs})$ is evaluated at $f_{obs}=0.1~{\rm yrs}^{-1}$. It is also convenient to present this limits in terms of the density parameter $\Omega_{gw}$
\bea
\Omega_{gw} \lesssim 5.3{\times}10^{-10}\left[\left(1-n_T/4\right)\left(\frac{R_{rms}}{0.1~{{\mu}\mathrm{sec}}}\right)^2\left(\frac{10~{\mathrm{yrs}}}{T_{obs}}\right)^4\right], ~~~~\textrm{for}~\epsilon\ll\epsilon_*
\label{omegalimit}
\ena
or
\bea
\Omega_{gw} \epsilon^2 \lesssim 4.0{\times}10^{-15}\left[\left(1-n_T/3\right) \left(\frac{10~{\mathrm{kpc}}}{D}\right) \left(\frac{R_{rms}}{0.1~{{\mu}\mathrm{sec}}}\right)^2\left(\frac{10~{\mathrm{yrs}}}{T_{obs}}\right)^3\right], ~~~~\textrm{for}~\epsilon\gg\epsilon_*
\label{omegaepsilonsquarelimit}
\ena
Thus, from (\ref{hc_limit}) (or (\ref{omegalimit})), it can be seen that for $\epsilon\ll\epsilon_*$, when surfing effect is not important, pulsar timing sets limits directly on $h_c$ (or equivalently on $\Omega_{gw}$), i.e.~the strength of the gravitational wave background. On the other hand, when $\epsilon\gg\epsilon_*$ and surfing effect becomes dominant, from (\ref{hcepsilon_limit}) (or (\ref{omegaepsilonsquarelimit})), it follows that pulsar timing sets limits on the combination $h_c\epsilon$ (or $\Omega_{gw}\epsilon^2$ equivalently). The upper limits from pulsar timing, along with possible sources and sensitivity levels of various experimental techniques to detect gravitational waves, are illustrated on figure \ref{figure2}.

As follows from the above discussion, and can be seen from figure \ref{figure2}, an independent knowledge of $h_c$ would enable us to directly constrain the parameter $\epsilon$, i.e.~constrain the deviation of speed of gravitational waves from speed of light. From expression (\ref{hcepsilon_limit}) we arrive at the following constrain on $\epsilon$
\bea
\epsilon \lesssim 1.1{\times}10^{-2}\left[\sqrt{1-2\alpha}\left(\frac{10^{-15}}{h_{c}}\right) \left(\frac{10~{\mathrm{kpc}}}{D}\right)^{\frac{1}{2}} \left(\frac{R_{rms}}{0.1~{{\mu}\mathrm{sec}}}\right) \left(\frac{10~{\mathrm{yrs}}}{T_{obs}}\right)^{\frac{3}{2}} \right].
\label{epsilonlimit2}
\ena
In terms of the density parameter $\Omega_{gw}$ the constraint has the form
\bea
\epsilon \lesssim 6.4{\times}10^{-3}\left[\sqrt{1-n_T/3}\left(\frac{10^{-10}}{\Omega_{gw}}\right)^{\frac{1}{2}} \left(\frac{10~{\mathrm{kpc}}}{D}\right)^{\frac{1}{2}} \left(\frac{R_{rms}}{0.1~{{\mu}\mathrm{sec}}}\right) \left(\frac{10~{\mathrm{yrs}}}{T_{obs}}\right)^{\frac{3}{2}} \right].
\label{epsilonlimit1}
\ena

In the next section we shall discuss the various viable candidates for a stochastic gravitational wave background and explicitly calculate the achievable limits on $\epsilon$. We shall also discuss the implications of the surfing effect for theories with massive gravitons.


\section{The physical implications of the surfing effect \label{physicalconsequences}}

The analysis in Section \ref{upperlimits} indicates that the
surfing effect in pulsar timing can yield interesting constraints
on $\epsilon$ parameter and consequently the mass of graviton in a
sufficiently strong gravitational wave background with $\Omega_{gw}\sim
10^{-10}$ (see (\ref{epsilonlimit1})). It is important
to note that this method is fundamentally limited by the value
$\epsilon_*$, which is currently about $3\times 10^{-3}$ (see
(\ref{epsilonstar})). Although an increase in the time of observation will improve overall precision, it will also increase the value of $\epsilon_*$, thus worsening the potential constraints on $\epsilon$. In future, the method can become more
sensitive with implementation of large radio telescopes like 
the Square Kilometer Array (SKA) (see \cite{Kramer2004} for
detailed discussion of SKA and its usage in pulsar astrophysics), which
would improve the limitations to $\epsilon_*\sim 10^{-3}$.
Furthermore, as seen from expression (\ref{epsilonlimit1}) (or
(\ref{epsilonlimit2})), increasing the pulsar timing accuracy (for
example, using pulsar timing ensembles \cite{Manchester2007}) can
reduce the limit down to the critical value $\epsilon_*$.

The gravitational wave background, at the frequency range of our interest (${f_{gw}} \lesssim 0.1 \mathrm{{\mathrm{yrs}}}^{-1}$), consists of contribution from a variety of well established astrophysical and cosmological sources \cite{glpps2001} as well as possible contribution from exotic remnants of early universe \cite{Maggiore2000}, \cite{Hogan2006}. The strongest contribution to the gravitational wave background, at these frequencies, come from the background of extragalactic coalescing supermassive binary black holes (SMBH) \cite{WyitheLoeb2003}, \cite{JaffeBacker2003}, \cite{Enoki2004}, \cite{Sesana2008}. For this reason, below in the subsection \ref{bhbackground}, we shall study the implications of the surfing effect for this background. Following this, in subsection \ref{darkmattergw}, we shall analyze the consequences of the limitations on $\epsilon$ for theories with massive gravitons.


\subsection{Gravitational wave background from extragalactic black holes\label{bhbackground}}

As was mentioned above, one of the strongest sources for a stochastic gravitational wave background at frequency range of our interest, ${f_{gw}}\sim T_{obs}^{-1}\approx 0.1~{\mathrm{yrs}}^{-1} $, comes from the extragalactic black hole binaries. Various groups have conducted a theoretical study on the strength of this background \cite{JaffeBacker2003}, \cite{WyitheLoeb2003}, \cite{Enoki2004}, \cite{Sesana2008}. There is a general consensus on the expected gravitational wave strain for this background
\bea
h_c(f) \approx 10^{-16}\left(\frac{f}{1\mu{\rm Hz}}\right)^{-\frac{2}{3}},
\label{bhstrain}
\ena
corresponding to the value for the density parameter
\bea
\Omega_{gw}(f)  \approx 2.4{\times}10^{-10}\left(\frac{f}{0.1~{\mathrm{yrs}}^{-1}}\right)^{\frac{2}{3}}.
\label{bhOmega}
\ena
The uncertainty surrounding this value of $h_c$ arises mainly due to the uncertainty in the galaxy merger rates as well as some other astrophysical factors. Taking into account these uncertainties, the amplitude lies in the interval $h_c(f = 1\mu {\rm Hz}) \approx 2.5\times10^{-17}- 4\times10^{-16} $  \cite{Sesana2008}.

The expected strain $h_c$ from the background of SMBH allows to place significant bounds on the $\epsilon$ parameter. Substituting expression (\ref{bhstrain}) into expression (\ref{epsilonlimit2}), and setting $\alpha=-2/3$, we arrive at the following limit on upper $\epsilon$
\bea
\epsilon \lesssim 3.7{\times} 10^{-3}\left[\left(\frac{10~{\mathrm{kpc}}}{D}\right)^{\frac{1}{2}} \left(\frac{R_{rms}}{0.1~{{\mu}\mathrm{sec}}}\right) \left(\frac{10~{\mathrm{yrs}}}{T_{obs}}\right)^{\frac{3}{2}} \right].
\ena
Thus, the stochastic gravitational wave background of extragalactic SMBH mergers can potentially place very stringent constraints, $\epsilon\lesssim 0.4\%$, on the speed of gravitational waves.


\subsection{Implications for theories with massive gravitons \label{darkmattergw}}

The phenomenological parameter $\epsilon$ is directly related to the mass of the graviton $m_g$
(see (\ref{mgravitondef})). It is convenient to rewrite expression (\ref{mgravitondef}) in the form
\bea
\epsilon(k) = \epsilon_o\left(\frac{k_{min}}{k}\right), ~~~~ \textrm{where}~ \epsilon_o = \frac{m_gc^2T_{obs}}{2\pi\hbar},
\label{epsilon_o}
\ena
For the fiducial strength of gravitational wave background we get
\bea
\epsilon_o \lesssim 8.3{\times}10^{-3}\left[\sqrt{1-n_T/5}\left(\frac{10^{-10}}{\Omega_{gw}}\right)^{\frac{1}{2}} \left(\frac{10~{\mathrm{kpc}}}{D}\right)^{\frac{1}{2}} \left(\frac{R_{rms}}{0.1~{{\mu}\mathrm{sec}}}\right) \left(\frac{10~{\mathrm{yrs}}}{T_{obs}}\right)^{\frac{3}{2}} \right].
\label{massivegravitonlimit1}
\ena
Note that the factor $n_T/5$, in the above expression (\ref{massivegravitonlimit1}) (compared with the factor $n_T/3$ in expression (\ref{epsilonlimit1})), arises because we are constraining $\epsilon_o$ (compared with constraints on $\epsilon$ in (\ref{epsilonlimit1})). This leads to an extra factor $\left(k/k_{min}\right)$ in integral $(\ref{Rsquaremean})$ and hence slightly modifies the result. The above limit on $\epsilon_o$ implies the following limit on the mass of the graviton
\bea
m_g \lesssim 1.1{\times}10^{-25}~\mathrm{eV}~\left[\sqrt{1-n_T/5}\left(\frac{10^{-10}}{\Omega_{gw}}\right)^{\frac{1}{2}} \left(\frac{10~{\mathrm{kpc}}}{D}\right)^{\frac{1}{2}} \left(\frac{R_{rms}}{0.1~{{\mu}\mathrm{sec}}}\right) \left(\frac{10~{\mathrm{yrs}}}{T_{obs}}\right)^{\frac{3}{2}} \right].
\label{massivegravitonlimit2}
\ena
From expression (\ref{epsilon_o}) it follows that, stronger constraint on $m_g$ require smaller values of $k_{min}$, i.e. require a longer time of observation $T_{obs}$. On the other hand, the strongest possible constraint for $\epsilon_o$ is determined by the value of $\epsilon_*$ (which increases with the time of observation, see expression (\ref{epsilonstar})). For this reason, an increase in $T_{obs}$ beyond a value of approximately $25~{\rm yrs}$ will not lead to an improvement in constraining $m_g$.

As a concrete example, let us assume that the gravitational wave background from SMBH coalesces dominates at frequencies $0.1-1~{\rm yrs}^{-1}$, and that its properties
are not affected by the non-zero mass of graviton. Then the existing four-years precise timing of PSR B1937+21 \cite{Manchester2007} allow to significantly constrain the mass of the graviton.
Setting $T_{obs}=4~{\rm yrs}$, $R_{rms}=0.17~{\mu}{\rm sec}$, $D=8.3~{\rm kpc}$, $n_T = 2/3$ and $\Omega_{gw}(T_{obs}^{-1}) = 4.2\times10^{-10}$ (see (\ref{bhOmega})) in expression (\ref{massivegravitonlimit2}), we arrive at a limit 
\bea 
m_g\lesssim 3.6\times10^{-25}~{\rm eV},
\ena 
corresponding to a Compton length for graviton of $\lambda_g={h}/{m_g c}\gtrsim3.4\times10^{15}\, {\rm km}$. This bound is three orders stronger than the current limit from Solar system tests \cite{Talmadge1988} and is comparable to future limits from SMBH mergers
obtainable with LISA (see \cite{Will2006} and references therein).
It is worth stressing, that the limits from pulsar timing are
more robust and less  model dependent than the prospects for LISA.

The surfing effect in pulsar timing puts stringent constraint on the mass of graviton in some theories of gravity (see \cite{ptp08}). In \cite{dtt2005} the authors propose massive gravitons as a viable candidates for cold dark matter in the galactic halo. At the frequency ranges of our interest, these massive gravitons imply $\epsilon\approx 0.5$. The existing precise timing of PSR B1937+21 place direct limits on the parameter  $\Omega_{gw}\epsilon^2\lesssim 2{\times}10^{-13}$ (setting $R_{rms}=0.17~{{\mu}\mathrm{sec}}$ and $T_{obs}=4~{\mathrm{yrs}}$ in expression (\ref{omegaepsilonsquarelimit})) . This implies that massive gravitons, as candidates to explain the dark matter in the galactic halo, can be ruled out with the current observations.


\section{Conclusions\label{conclusions}}

In this work we have analyzed the consequences of the surfing effect, introduced in \cite{PolnarevBaskaran2008}, for pulsar timing observations.  The surfing effect, due to the transverse nature of gravitational waves, leads to a strong observable signature only when the speed of gravitational waves is smaller than the speed of light. In order to analyze this possibility, we have introduced a parameter $\epsilon$, which characterizes the deviation of speed of gravitational waves from speed of light. By studying the pulsar timing residuals in the presence of a single plane monochromatic gravitational wave, followed by a generalization to an arbitrary gravitational wave field, we show the presence and importance of surfing effect in the case when $\epsilon\neq0$.

The surfing effect allows to place significant bounds on the parameter $\epsilon$. For a timing accuracy of $R_{rms}=0.1~{{\mu}\mathrm{sec}}$, and assuming a realistic background of gravitational waves from extragalactic super massive black hole binary mergers, the achievable limits are $\epsilon\lesssim 0.4\%$. The strongest achievable bounds on $\epsilon$ are determined by $\epsilon_*$. For a pulsar at a typical distance $D=10~{\mathrm{kpc}}$ the value is $\epsilon_*\approx 0.3\%$. This limit could potentially be slightly improved by observing pulsars at a greater distance $D$.

The surfing effect leads to interesting consequences for theories with massive gravitons. Using the existing observations, we have constrained the mass of graviton to $m_g\lesssim 4\times10^{-25}\, {\rm eV}$, which is  three orders of magnitude stronger than the current limits from Solar system tests. With future observations this constraint could improve by an order of magnitude. Based on the existing observations, we have also ruled out massive gravitons as candidates to explain the dark matter in the galactic halo.

In comparison with precision interferometry methods considered in \cite{PolnarevBaskaran2008}, pulsar timing measurements (due to their high precision) should be able to put tighter constraints on $\epsilon$. In any case, these two methods of constraining $\epsilon$ are independent and hence should be considered complementary.

%

\section*{Acknowledgements}
Authors thank B.~G.~Keating, W.~Zhao, L.~P.~Grishchuk and M.V.~Sazhin for useful discussions and fruitful suggestions. The work of MP is supported by RFBR grant 06-02-16816-a. KP acknowledges partial support from RFBR grant 07-02-00961-a.



\appendix

\section{Evaluation of the transfer function\label{AppendixA}}

Let us evaluate the integral in expression
(\ref{transferfunction2})
\bea
I(k)= \int\limits_{-1}^{+1} d\mu \left(1-\mu^2\right)^2 \left[ \frac{\sin^2\left\{\frac{kD}{2}\left(1-\epsilon-\mu\right) \right\}}{\left(1-\epsilon-\mu\right)^2} \right],
\label{SurfingIntegral}
\ena
in the physically interesting case when
$\epsilon\rightarrow 0$ and $kD\gg1$. The integral can be separated into two distinctive
contributions
\bea
I(k) = I_{NR}(k) +
I_{R}(k),
\ena
where $I_{NR}(k)$ is the non-resonance
contribution
\bea
I_{NR}(k) & = &
\int\limits_{-1}^{1-\epsilon-\Delta\mu} d\mu ~ \left(
1-\mu^2 \right)^2 \left[\frac{\sin^2{\left\{ \frac{\pi
D}{\lambda_{gw}} \left(1 - \epsilon - \mu\right)
\right\}}}{\left(1 - \epsilon - \mu\right)^2}\right] \nonumber \\
& & + \int\limits_{1-\epsilon+\Delta\mu}^{+1} d\mu ~
\left( 1-\mu^2 \right)^2 \left[\frac{\sin^2{\left\{ \frac{\pi
D}{\lambda_{gw}} \left(1 - \epsilon - \mu\right)
\right\}}}{\left(1 - \epsilon - \mu\right)^2}\right],
\label{NRintegral}
\ena
and $\Delta\tilde{\alpha}_{R}^2(k)$ is the resonance (or, in other words ``surfing",) contribution
\bea
\Delta\tilde{\alpha}_{R}^2(k) & = &
\int\limits_{1-\epsilon-\Delta\mu}^{1-\epsilon+\Delta\mu}
d\mu ~ \left( 1-\mu^2 \right)^2 \left[\frac{\sin^2{\left\{
\frac{\pi D}{\lambda_{gw}} \left(1 - \epsilon - \mu\right)
\right\}}}{\left(1 - \epsilon - \mu\right)^2}\right].
\label{Rintegral}
\ena
The quantity $\Delta\mu$ occurring in the limits of integration in
the above expressions is fixed by the condition for the resonance
to occur. This condition corresponds to the region, around
$\mu=1-\epsilon$, where the sine function undergoes a few
oscillations. Thus $\Delta\mu = N\lambda_{gw}/D = 2\pi N/kD$,
where $N$ is the number of oscillations of the sine function,
around the point $\mu=1-\epsilon$, included in evaluation of the
resonance. The value of $N$ is limited by the condition $\Delta\mu
= 2\pi N/kD\ll\epsilon$, implying $N\ll\epsilon kD/2\pi$. Since in
all our considerations we assume $\epsilon\ll 1$, and
$\epsilon^2kD\gg1$, the condition imposed on $N$ is consistent
with an additional condition $N\gg1$ that we shall assume.

When evaluating (\ref{NRintegral}), since we assume
$\epsilon\ll 1$, we can neglect the second integral in
comparison with the first. In evaluation fo the remaining integral
we can set $\epsilon = 0$. Thus, we get
\bea
I_{NR}(k) & \approx &
\int\limits_{-1}^{1} d\mu ~\left(
1+\mu \right)^2 \sin^2{\left( \frac{k D}{2} \left(1 - \mu\right)
\right)} \nonumber \\ & =& \frac{1}{2}\int\limits_{-1}^{1} d\mu ~
\left( 1+\mu \right)^2
\left(1-\frac{}{}\cos{\left( {k D} \left(1 -
\mu\right)\right)}\right)\nonumber \\ & \approx &
\frac{1}{2}\int\limits_{-1}^{1} d\mu ~ \left(
1+\mu \right)^2 = \frac{4}{3},\label{NRintegral2}
\ena
where, assuming $kD\gg1$, we have explicitly separate out the
rapid oscillatory part and neglected it in the last line.

In order to evaluate (\ref{Rintegral}), in the case of
$\epsilon\ll 0$ and $kD\gg1$, it is helpful to notice that
the factor $\left( 1-\mu^2 \right)^2$ in the right side of
(\ref{Rintegral}) is a slowly varying function over the range of
integration. Taking this factor (evaluated at $\mu=1-\epsilon$)
outside the integral we get the following approximation for the
resonance part of the transfer function
\bea
I_{R}(k) & \approx &
4\epsilon^2\int\limits_{1-\epsilon-\Delta\mu}^{1-\epsilon+\Delta\mu}
d\mu ~ \left[\frac{\sin^2{\left\{ \frac{k D}{2} \left(1 - \epsilon
- \mu\right) \right\}}}{\left(1 - \epsilon - \mu\right)^2}\right]
 =
2\epsilon^2kD\int\limits_{-N\pi}^{+N\pi}dx\frac{\sin^2{x}}{x^2}\nonumber \\
~&\approx& ~
2\pi\epsilon^2kD\left(1-O\left(\frac{1}{N^2}\right)\right)
~ \approx ~ 2\pi\epsilon^2kD.
\label{Rintegral2}
\ena

Finally, the total transfer function, given by the sum of the
non-resonance (\ref{NRintegral2}) and resonance parts
(\ref{Rintegral2}), has the following form
\bea
I(k) = I_{NR}(k)+I_R(k) \approx \frac{4}{3}\left[1 +\frac{3}{2}
\pi\epsilon^2kD\right]. \label{AppendixAintegralTotal}
\ena


\end{document}